\documentclass[aps,pra,10pt,twocolumn,groupedaddress,nofootinbib, longbibliography]{revtex4-1}

\usepackage{amsmath, amssymb, amsthm}
\usepackage{graphicx}
\usepackage{bm,bbm}
\usepackage{hyperref}
\usepackage[margin=2cm]{geometry}
\usepackage{xcolor}
\usepackage{algorithm}
\usepackage[noend]{algpseudocode}
\usepackage{qcircuit}
\usepackage{multirow}
\theoremstyle{plain}
\usepackage{tikz}
\usepackage{url}
\newtheorem{theorem}{Theorem}
\usepackage{mathtools}

\newcommand{\hc}{\text{h.c.}}  
\newcommand{\bpm}{\begin{pmatrix}}
\newcommand{\epm}{\end{pmatrix}}

\newcommand{\R}{\ensuremath{\mathbb R}}  
\newcommand{\I}{\mathbbm{1}} 
\newcommand{\cell}[2][c]{\begin{tabular}[#1]{@{}l@{}}#2\end{tabular}}

\newcommand{\ket}[1]{\ensuremath{\left| #1 \right \rangle}}
\newcommand{\bra}[1]{\ensuremath{\left \langle #1 \right |}}

\newcommand{\ketbra}[2]{\ket{#1}\bra{#2}}
\newcommand{\expval}[1]{\ensuremath{\langle #1 \rangle}}            

\newcommand{\x}{\hat{x}}
\newcommand{\p}{\hat{p}}

\newcommand{\e}{\mathrm{e}}

\newcommand{\G}{\mathcal{G}}

\newcommand{\vc}{f}  
\newcommand{\func}{g}  

\definecolor{darkgreen}{HTML}{418356}

\usetikzlibrary{calc}
\usetikzlibrary{shapes.multipart}
\definecolor{quantum1}{HTML}{8EDBCE}
\definecolor{quantum2}{HTML}{3F605B}
\definecolor{exp1}{HTML}{9AB9ED}
\definecolor{exp2}{HTML}{204177}
\definecolor{classical}{HTML}{EBBA92}
\tikzset{input node/.style={}}
\tikzset{quantum node/.style={draw, align=center, anchor=west, inner sep=5pt,rounded corners=4pt, rectangle split, rectangle split horizontal, rectangle split parts=2, rectangle split part fill={quantum1,quantum2}, every two node part/.style={text = white}}}
\tikzset{classical node/.style={draw, rectangle,align=center, anchor=west, thin, fill=classical, inner sep=5pt}}
\tikzset{output node/.style={}}
\tikzset{out label/.style={midway, above}}
\tikzset{connector/.style={anchor=center, opacity=0.}}
\tikzset{samples label/.style={at start, below, xshift=4pt}}

\begin{document}

\title{Evaluating analytic gradients on quantum hardware}
\author{Maria Schuld}
\email{maria@xanadu.ai}
\author{Ville Bergholm}
\author{Christian Gogolin}
\author{Josh Izaac}
\author{Nathan Killoran}

\affiliation{Xanadu Inc., 372 Richmond St W, Toronto, M5V 1X6, Canada}
\date{\today}

\begin{abstract}
An important application for near-term quantum computing lies in optimization tasks,
with applications ranging from quantum chemistry and drug discovery to machine learning.
In many settings --- most prominently in so-called parametrized or variational algorithms ---
the objective function is a result of hybrid quantum-classical processing.
To optimize the objective, it is useful to have access to exact gradients of quantum circuits with respect to gate parameters.
This paper shows how gradients of expectation values of quantum measurements can be estimated
using the same, or almost the same, architecture that executes the original circuit.
It generalizes previous results for qubit-based platforms, and proposes
recipes for the computation of gradients of continuous-variable circuits.
Interestingly, in many important instances it is sufficient to run the original quantum circuit twice while shifting a single gate parameter to obtain the corresponding component of the gradient.
More general cases can be solved by conditioning a single gate on an ancilla.
\end{abstract}

\maketitle

\section{Introduction}

\begin{figure*}[t]
\centering
\includegraphics[width=0.9\textwidth]{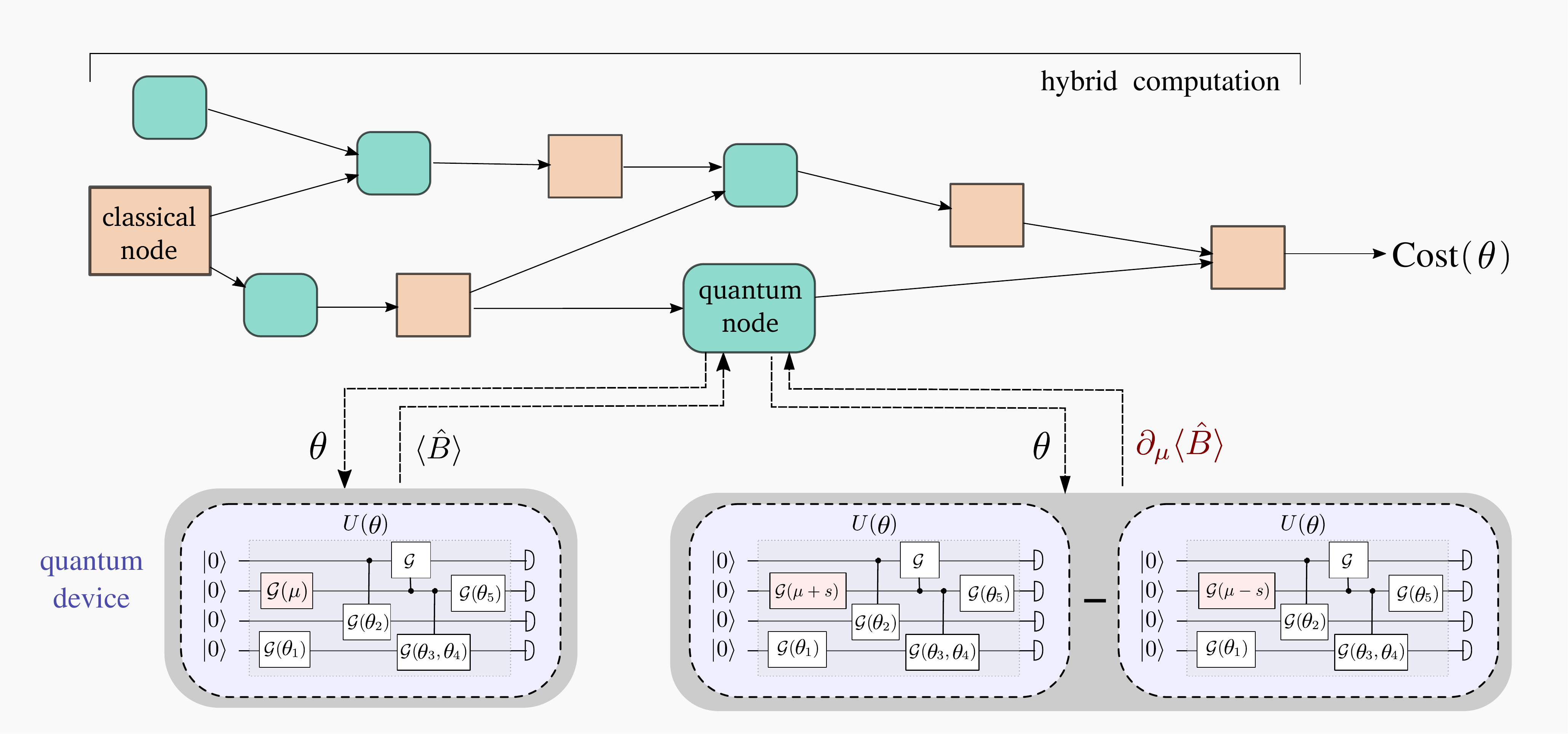}
\caption{The ``parameter shift rule'' in the larger context of hybrid optimization.
  A quantum node, in which a variational quantum algorithm is executed, can compute derivatives of its outputs with respect to gate parameters by running the original circuit twice, but with a shift in the parameter in question.}
\label{Fig:hybrid}
\end{figure*}

Hybrid optimization algorithms have become a central quantum software design paradigm for current-day quantum technologies, since they outsource parts of the computation to classical computers.
Examples of such algorithms are variational quantum eigensolvers \cite{peruzzo2014variational}, quantum approximate optimization \cite{farhi2014quantum}, variational autoencoders \cite{romero2017quantum}, quantum feature embeddings \cite{schuld2018quantum, havlicek2018supervised} and variational classifiers \cite{schuld2018circuit, farhi2018classification}, but also more general hybrid optimization frameworks \cite{bergholm2018pennylane}.
In such applications, the objective or cost function is a combination of both classical and quantum information processing modules, or nodes (see Fig.~\ref{Fig:hybrid}).
The quantum nodes execute parametrized quantum circuits, also called \textit{variational circuits}, in which gates have adjustable continuous parameters such as rotations by an angle.

To unlock the potential of gradient-descent-based optimization strategies it is essential to have access to the gradients of quantum computations.
While individual quantum measurements produce probabilistic results, the expectation value of a quantum observable --- which can be estimated by taking the average over measurement results --- is a deterministic quantity that varies smoothly with the gate parameters.
It is therefore possible to formally define the gradient of a quantum computation via derivatives of expectations.

The challenge however is to compute such gradients on quantum hardware.
As we will lay out below, the derivative of a quantum expectation with respect to a parameter $\mu$ used in gate $\G$ involves the ``derivative of the gate'' $\partial_{\mu} \G$, which is not necessarily a quantum gate itself. Hence, the derivative of an expectation is not a valid quantum expectation.
Since in interesting cases the gradient, just as the objective function itself, tends to be classically intractable, we need to express such derivatives as a combination of quantum operations that can be implemented in hardware.
Even more, in the case of special-purpose quantum hardware it is desirable that gradients can be evaluated by the same device that is used for the original computation.

This paper derives rules to compute the partial derivatives of quantum expectation values with respect to gate parameters on quantum hardware.
A number of results in this direction have been recently proposed in the quantum machine learning literature \cite{guerreschi2017practical, farhi2018classification, schuld2018circuit, mitarai2018quantum, liu2018differentiable}.
Refs.~\cite{guerreschi2017practical, farhi2018classification, schuld2018circuit} note that if the derivative $\partial_{\mu} \G$ as well as the observable whose expectation we are
interested in can be decomposed into a sum of unitaries, we can evaluate the derivative of an expectation by measuring an overlap of two quantum states.
Mitarai \emph{et al.}~\cite{mitarai2018quantum}, leveraging a technique from quantum control, propose an elegant method for gates of the form
$\G = e^{-i \mu \sigma}$, where $\sigma$ is a tensor product of the Pauli operators $\{\sigma_x, \sigma_y, \sigma_z\}$.
In this case, the derivative can be computed by what we will call the ``parameter shift rule'', which requires us to evaluate the original expectation twice, but with one circuit parameter shifted by a fixed value.

In this work, we make several contributions to the literature on quantum gradients.
Firstly, we expand the parameter shift rule by noting that it holds for any gate
of the form $\G = e^{-i \mu G}$, where the Hermitian generator~$G$ has at most two distinct eigenvalues.
We mention important examples of this class.
Secondly, we show that any other gate can be handled by a method that involves a coherent \textit{linear combination of unitaries} routine \cite{childs12}. This requires adding a single ancilla qubit and conditioning the gate and its ``derivative'' on the ancilla while running the circuit.
Thirdly, we derive parameter shift rules for Gaussian gates in continuous-variable quantum computing. These rules can be efficiently implemented if all gates following the differentiated gate are Gaussian and the final observable is a low-degree polynomial of the creation and annihilation operators. In fact, the method still works efficiently for some non-Gaussian gates, such as the cubic phase gate, as long as there is at most a logarithmically large number of these non-Gaussian gates.
The results of this paper are implemented in the software framework \textit{PennyLane} \cite{bergholm2018pennylane}, which facilitates hybrid quantum-classical optimization across various quantum hardwares and simulator platforms \cite{bergholm2018pennylane}.

\section{Computing quantum gradients}

Consider a quantum algorithm that is possibly part of a larger hybrid computation, as shown in Fig.~\ref{Fig:hybrid}.
The quantum algorithm or circuit consists of a gate sequence $U(\theta)$ that depends on a set $\theta$ of $m$ real gate parameters,
followed by the measurement of an observable~$\hat{B}$.\footnote{
  The output of the circuit may consist of the measurements of $n$~mutually commuting scalar observables,
  however, without loss of generality, they can always be combined into a vector-valued observable with $n$~components.
}
An example is the Pauli-Z observable $\hat{B} = \sigma_z$, and the result of this single measurement is $\pm 1$
for a qubit found in the state $\ket{0}$ or $\ket{1}$, respectively.
The gate sequence $U(\theta)$ usually consists of an ansatz or architecture that is repeated $K$ times, where $K$ is a hyperparameter of the computation.

We refer to the combined procedure of applying the gate sequence~$U(\theta)$ and finding the
expectation value of the measurement~$\hat{B}$ as a \textit{variational circuit}.
In the overall hybrid computation one can therefore understand a variational circuit as a function~$\vc: \R^m \to \R^n$, mapping the gate parameters to an expectation,
\begin{equation}
\vc(\theta) \coloneqq \expval{\hat{B}} = \bra{0} U^{\dagger}(\theta) \hat{B} U(\theta) \ket{0}.
\label{Eq:expval}
\end{equation}
While this abstract definition of a variational circuit is exact, its physical implementation on a quantum device runs the quantum algorithm several times and averages measurement samples to get an \textit{estimate} of~$\vc(\theta)$.
If the circuit is executed on a classical simulator, $\vc(\theta)$ can be computed exactly up to numerical precision.

In the following, we are concerned with the partial derivative
$\partial_{\mu} \vc(\theta)$
where $\mu \in \theta$ is one of the gate parameters.
The partial derivatives with respect to all gate parameters form the gradient $\nabla \vc$.
The differentiation rules we derive consider the expectation value in Eq.~\eqref{Eq:expval} and are therefore exact.
Just like the variational circuit itself has an `analytic' definition and a `stochastic' implementation,
the \textit{evaluation} of these rules with finite runs on noisy hardware return estimates of the gradient.\footnote{It is an open question whether such estimates have favourable properties similar to approximations of gradients in stochastic gradient descent.}

There are three main approaches to evaluate the gradients of a numerical computation,
i.e., a computer program that executes a mathematical function $\func(x)$:
\begin{enumerate}
\item \textit{Numerical differentiation}: The gradient is approximated by black-box evaluations of $\func$, e.g.,
  \begin{equation}
    \label{eq:finitediff}
    \quad \qquad \nabla \func(x) \approx (\func(x +\Delta x/2) -\func(x -\Delta x/2) )/\Delta x,
  \end{equation}
  where $\Delta x$ is a small shift.
\item \textit{Automatic differentiation}: The gradient is efficiently computed through the accumulation of intermediate derivatives
  corresponding to different subfunctions used to build $\func$, following the chain rule~\cite{maclaurin2015autograd}.
\item \textit{Symbolic differentiation}: Using manual calculations or a symbolic computer algebra package,
  the function $\nabla \func$ is constructed and evaluated.
\end{enumerate}
Until recently, numerical differentiation (or altogether gradient-free methods) have been the
method of choice in the quantum variational circuits literature.
However, the high errors of near-term quantum devices can make it unfeasible
to use finite difference formulas to approximate the gradient of a circuit.

Several modern numerical programming frameworks, especially in machine learning, successfully employ automatic differentiation \cite{baydin2017automatic} instead, a famous example being the ubiquitous backpropagation algorithm for the training of neural networks.
Unfortunately, it is not clear how intermediate derivatives could be stored and reused \emph{inside of a quantum computation}, since the intermediate quantum states cannot be measured without impacting the overall computation.

To compute gradients of quantum expectation values, we therefore use the following strategy:
Derive an equation for $\partial_{\mu} \vc(\theta), \;\mu \in \theta$, whose
constituent parts can be evaluated on a quantum computer and subsequently combined on a classical coprocessor.
It turns out that this strategy has a number of favourable properties:
\begin{enumerate}
\item It follows similar rules for a range of different circuits,
\item Evaluating $\partial_{\mu} \vc(\theta)$ can often be done on a circuit architecture that is very similar or even identical to that for evaluating $\vc(\theta)$,
\item Evaluating $\partial_{\mu} \vc(\theta)$ requires the  evaluation of only two expectation values.
\end{enumerate}
We emphasize that automatic differentiation techniques such as backpropagation can still be used within a larger overall hybrid computation, but we will not get any efficiency gains for this technique on the intermediate steps of the quantum circuit.

The remainder of the paper will present the recipes for how to evaluate the derivatives of expectation values,
first for qubit-based, and then for continuous-variable quantum computing.
The results are summarized in Table~\ref{Tbl:results}.

\begin{table*}[t]
\def\arraystretch{1.5}
\setlength\tabcolsep{12pt}
\begin{tabular}{p{4cm} p{6cm} p{4.5cm}}
\hline \hline
Architecture & Condition & Technique \\ \hline
Qubit & $\G$ generated by a Hermitian operator with $2$ unique eigenvalues &  parameter shift rule\\
Qubit & no special condition & derivative gate decomposition  + linear combination of unitaries\\
Continuous-variable & $\G$ Gaussian, followed by at most logarithmically many non-Gaussian operations & continuous-variable parameter shift rules \\
Continuous-variable & no special condition & unknown \\ \hline \hline
\end{tabular}
\caption{Summary of results. $\G$ refers to the gate with parameter $\mu$ that we compute the partial derivative for. $\partial_{\mu} \G$ refers to the partial derivative of the operator $\G$. }
\label{Tbl:results}
\end{table*}


\section{Gradients of discrete-variable circuits}
\label{sec:discrete}
As a first step, the overall unitary
$U(\theta)$ of the variational circuit can be decomposed into a sequence of single-parameter gates,
which can be differentiated using the product rule.
For simplicity, let us assume that the parameter $\mu \in \theta$ only affects
a single gate $\G(\mu)$ in the sequence, $U(\theta) = V \G(\mu) W$.
The partial derivative $\partial_\mu \vc$ then looks like
\begin{equation}
  \label{Eq:der_of_exp}
  \partial_\mu \vc = \partial_{\mu} \bra{\psi} \G^{\dagger} \hat{Q} \G\ket{\psi}
  = \bra{\psi} \G^{\dagger}  \hat{Q}  (\partial_{\mu}\G) \ket{\psi} +\hc,
\end{equation}
where we have absorbed~$V$ into the Hermitian observable $\hat{Q} = V^\dagger \hat{B} V$,
and $W$~into the state $\ket{\psi} = W \ket{0}$.

For any two operators $B$, $C$ we have
\begin{align}
  \label{Eq:term_equivalence}
  \bra{\psi}B^{\dagger} \hat{Q} C  \ket{\psi} +\hc\notag\\
  \begin{split}
    = \frac{1}{2}\big(&
    \bra{\psi} (B + C)^{\dagger} \hat{Q} (B + C) \ket{\psi}\\
    -&\bra{\psi}(B - C)^{\dagger}  \hat{Q} (B - C) \ket{\psi}
  \big).
\end{split}
\end{align}
Hence, whenever we can implement $\G \pm \partial_\mu \G$ as part of an overall unitary evolution, we can evaluate
Eq.~\eqref{Eq:der_of_exp} directly.
Sec.~\ref{sec:parameter_shift_rule} identifies a class of gates for which $\G \pm \partial_\mu \G$ is already unitary, while Sec.~ \ref{sec:qubit_general_lcu} shows that an ancilla can help to evaluate the terms in Eq.~\eqref{Eq:der_of_exp} with minimal overhead, and guaranteed success.

\subsection{Parameter-shift rule for gates with generators with two distinct eigenvalues}
\label{sec:parameter_shift_rule}
Consider a gate $\G(\mu) = \e^{-i \mu G}$ generated by a Hermitian operator~$G$.
Its derivative is given by
\begin{equation}
 \partial_{\mu} \G = -i G \e^{-i \mu G}.
\end{equation}
Substituting into Eq.~\eqref{Eq:der_of_exp}, we get
\begin{align}
  \partial_{\mu} \vc
   &= \bra{\psi'}  \hat{Q} \, (-iG) \ket{\psi'} +\hc ,
\label{Eq:qubitder}
\end{align}
where $\ket{\psi'} = \G \ket{\psi}$.
If $G$ has just two distinct eigenvalues (which can be repeated)
\footnote{The rather elegant special case for generators $G$
  that are tensor products of Pauli matrices has been presented in Mitarai \emph{et al.}~\cite{mitarai2018quantum}.
  Here we consider the slightly more general case.}
we can, without loss of generality, shift the eigenvalues to~$\pm r$, as the global phase is unobservable.
Note that any single qubit gate is of this form.
Using Eq.~\eqref{Eq:term_equivalence} for $B=\I$ and $C=-ir^{-1}G$ we can write
\begin{equation}
  \begin{split}
    \partial_{\mu} \vc = \frac{r}{2}\big(
    &\bra{\psi'} (\I -ir^{-1}G)^{\dagger} \hat{Q} (\I -ir^{-1}G) \ket{\psi'}\\
    -&\bra{\psi'}(\I +ir^{-1}G)^{\dagger}  \hat{Q} (\I +ir^{-1}G) \ket{\psi'}
    \big).
  \end{split}
\end{equation}
We now show that for gates with eigenvalues $\pm r$ there exist values for $\mu$ for which $\G(\mu)$ becomes equal to $\frac{1}{\sqrt{2}}(\I \pm i r^{-1}G)$.
\begin{theorem} 
\label{Th:taylor}
If the Hermitian generator~$G$ of the unitary operator $\G(\mu) = \e^{-i \mu G}$
has at most two unique eigenvalues $\pm r$,
the following identity holds:
\begin{equation}
  \G\left(\frac{\pi}{4r}\right) = \frac{1}{\sqrt{2}}(\I - i r^{-1} G).
  \label{Eq:Pauli_equality}
\end{equation}
\begin{proof}
The fact that $G$ has the spectrum $\{\pm r\}$ implies $G^2 = r^2 \I$.
Therefore the sine and cosine parts of the Taylor series of $\G(\mu)$ take the following simple form:
\begin{align}
\label{Eq:taylor}
\G(\mu) &= \exp(-i\mu G) = \sum_{k=0}^\infty \frac{(-i\mu)^k G^k}{k!}\\
&=
\sum_{k=0}^\infty \frac{(-i\mu)^{2k} G^{2k}}{(2k)!}
+\sum_{k=0}^\infty \frac{(-i\mu)^{2k+1} G^{2k+1}}{(2k+1)!} \\
&=
\I \sum_{k=0}^\infty \frac{(-1)^{k} (r\mu)^{2k}}{(2k)!} \nonumber \\
& \quad -i r^{-1} G \sum_{k=0}^\infty \frac{(-1)^{k} (r\mu)^{2k+1}}{(2k+1)!}\\
&=
\I \cos(r \mu)
-i r^{-1} G \sin(r \mu).
\end{align}
Hence we get
$\G(\frac{\pi}{4r}) = \frac{1}{\sqrt{2}}(\I -i r^{-1} G)$.
\end{proof}
\end{theorem}

We conclude that in this case $\partial_{\mu} \vc$
can be estimated using two additional evaluations of the quantum device;
for these evaluations, we place either the gate $\G(\frac{\pi}{4r})$ or the gate
$\G(-\frac{\pi}{4r})$ in the original circuit next to the gate we are differentiating.
Since for unitarily generated one-parameter gates $\G(a) \G(b) = \G(a+b)$,
this is equivalent to shifting the gate parameter, and we get the ``parameter shift rule'' with the shift $s = \frac{\pi}{4r}$:
\begin{align}
\label{eq:parameter_shift_rule}
\partial_{\mu} \vc
&=
r \big(\bra{\psi} \G^\dagger(\mu +s) \hat{Q} \G(\mu +s) \ket{\psi}\\
\notag
& \quad -\bra{\psi} \G^{\dagger}(\mu -s) \hat{Q}  \G(\mu -s) \ket{\psi}\big)\\
\label{eq:parameter_shift_rule2}
&= r \left(f(\mu+s) -f(\mu-s)\right).
\end{align}

If the parameter $\mu$ appears in more than a single gate in the circuit,
the derivative is obtained using the product rule by shifting the parameter in each gate separately and summing the results.
It is interesting to note that Eq.~\eqref{eq:parameter_shift_rule2} looks similar to the finite difference rule
in Eq.~\eqref{eq:finitediff}, but uses a macroscopic shift and is in fact exact.

The parameter shift rule applies to a number of special cases. As remarked in Mitarai \emph{et al.} \cite{mitarai2018quantum}, if $G$ is a one-qubit rotation generator in $\frac{1}{2}\{\sigma_x, \sigma_y, \sigma_z\}$ then
$r=1/2$ and $s = \frac{\pi}{2}$.
If $G = r \vec{n} \cdot\boldsymbol{\sigma}$ is a linear combination of Pauli operators with the $3$-dimensional normal vector $\vec{n}$, it still has two unique eigenvalues and Eq.~\eqref{Eq:Pauli_equality} can also be derived from what is known as the generalized Euler rule.

Also gates from a ``hardware-efficient'' variational circuit ansatz may fall within the scope of the parameter shift rule.
For example, according to the documentation of Google's \textit{Cirq} programming language \cite{googlecirq}, their Xmon qubits naturally implement the three gates
\begin{align*}
  \text{ExpW}(\mu, \delta) &= \exp \left(- i \mu \left( \cos (\delta) \sigma_x + \sin(\delta)\sigma_y\right) \right),\\
  \text{ExpZ}(\mu) &= \exp \left(- i \mu \sigma_z \right),\\
  \text{Exp11}(\mu) &= \exp \left(- i \mu \ketbra{11}{11}  \right) .
\end{align*}
which all have generators with at most two eigenvalues.

Pauli-based multi-qubit gates however do in general not fall in this category. A hardware-efficient example here is the microwave-controlled transmon gate for superconducting architectures\footnote{The time-dependent prefactors are summarized as the gate parameter $\mu$, while $b=\frac{J}{\Delta_{12}}$ represents the quotient of the interaction strength $J$ and detuning $\Delta_{12}$ between the qubits, and $c =m_{12}$ is a cross-talk factor.} \cite{chow2011simple},
\begin{equation*}
  \G(\mu) = \exp \left( \mu (\sigma_x \otimes \I -  b (\sigma_z \otimes \sigma_x) + c (\I \otimes \sigma_x ))  \right).
\end{equation*}
which has $4$ eigenvalues.
In these cases, other strategies have to be found to compute exact gradients of variational circuits.

\subsection{Differentiation of general gates via linear combination of unitaries}
\label{sec:qubit_general_lcu}

\begin{figure}[t]
$$
\Qcircuit @C=1em @R=.7em {
\lstick{\ket{0}} &  \gate{H}  &\ctrlo{1} & \ctrl{1}&  \gate{H} & \meter & \cw & \rstick{c}\\
\lstick{\ket{\psi}} &   \qw &\gate{\G} & \gate{A} & \qw  & \qw & \qw & \rstick{\ket{\psi'}}  \\
}
$$
\caption{Quantum circuit illustrating the `linear combination of unitaries' technique \cite{childs12}.
  Between interfering Hadamards, two unitary circuits or gates $A$ and~$\G$ are applied conditioned on an ancilla.
  Depending on the state of the ancilla qubit, the effect is equivalent to applying a sum or difference of $A$ and~$\G$.
\label{Fig:lcu_qubits}
}
\end{figure}
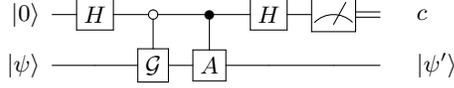

In case the parameter-shift differentiation strategy does not apply,
we may always evaluate Eq.~\eqref{Eq:der_of_exp} by introducing an ancilla qubit.
Since for finite-dimensional systems $\partial_{\mu}\G$ can be expressed as a complex square matrix,
we can always decompose it into a linear combination of unitary matrices $A_1$ and $A_2$,
\begin{equation}
  \partial_{\mu}\G = \frac{\alpha}{2} ((A_1 + A_1^{\dagger})  + i(A_2 + A_2^{\dagger}))
  \label{Eq:decomp}
\end{equation}
with real $\alpha$.\footnote{
  If $\alpha$ contains a renormalisation so that $|\G| \leq 1$, and
  $\G = \G_{\mathrm{re}} + i \G_{\mathrm{im}}$ we can set
  $A_1 = \G_{\mathrm{re}} + i \sqrt{\I - \G_{\mathrm{re}}^2}$ and
  $A_2 = \G_{\mathrm{im}} + i \sqrt{\I - \G_{\mathrm{im}}^2}$.}
$A_1$ and $A_2$ in turn can be implemented as quantum circuits.
To be more general, for example when another decomposition suits the hardware better, we can write
\begin{equation}
  \partial_{\mu}\G = \sum_{k=1}^K \alpha_k A_k,
  \label{Eq:deriv_decomp}
\end{equation}
for real $\alpha_k$ and unitary $A_k$. The derivative becomes
\begin{equation}
  \partial_{\mu} \vc  = \sum_{k=1}^K \alpha_k  \left( \bra{\psi}\G^{\dagger} \hat{Q} A_k  \ket{\psi} +\hc \right).
  \label{Eq:gradient_decomp}
\end{equation}
With Eq.~\eqref{Eq:term_equivalence}
we may compute the value of each term in the sum
using a coherent linear combination of the unitaries $\G$ and~$A_k = A$,
implemented by the quantum circuit in Fig.~\ref{Fig:lcu_qubits} (here and in the following we drop the subscript $k$ for readability).

First, we append an ancilla in state $\ket{0}$ and apply a Hadamard gate to it to obtain the bipartite state
\begin{equation}
  \frac{1}{\sqrt{2}} \left( \ket{0} + \ket{1}  \right)\otimes \ket{\psi} .
\end{equation}
Next, we apply $\G$ conditioned on the ancilla being in state $0$, and $A$ conditioned on the ancilla being in state $1$ (remember that both $\G$ and $A$ are unitary). This results in the state
\begin{equation}
 \frac{1}{\sqrt{2}} \left( \ket{0} \G \ket{\psi} + \ket{1} A \ket{\psi} \right).
\end{equation}
Applying a second Hadamard on the ancilla we can prepare the final state
\begin{equation}
 \frac{1}{2} \left( \ket{0} (\G + A) \ket{\psi} + \ket{1} (\G - A) \ket{\psi} \right).
\end{equation}
A measurement of the ancilla selects one of the two branches and results in either the state
$\ket{\psi'_0} = \frac{1}{2 \sqrt{p_0}} (\G + A) \ket{\psi}$ with probability
\begin{equation}
p_0  = \frac{1}{4} \bra{\psi} (\G + A)^{\dagger} (\G + A) \ket{\psi},
\end{equation}
or the state $\ket{\psi'_1} = \frac{1}{2\sqrt{p_1}} (\G - A) \ket{\psi}$ with probability
\begin{equation}
p_1 = \frac{1}{4} \bra{\psi} (\G - A)^{\dagger} (\G - A) \ket{\psi}.
\end{equation}
We then measure the observable $\hat{Q}$ for the final state~$\ket{\psi'_i}$, $i=0,1$.
Repeating this process several times allows us to estimate $p_0$, $p_1$ and the expected values
of~$\hat{Q}$ conditioned on the value of the ancilla,
\begin{equation}
\tilde{E}_0 = \bra{\psi'_0} \hat{Q} \ket{\psi'_0} = \frac{1}{4 p_0} \bra{\psi} (\G + A)^{\dagger} \hat{Q} (\G + A) \ket{\psi},
\end{equation}
and
\begin{equation}
\tilde{E}_1 = \bra{\psi'_1} \hat{Q} \ket{\psi'_1} = \frac{1}{4 p_1} \bra{\psi} (\G - A)^{\dagger} \hat{Q} (\G - A) \ket{\psi}.
\end{equation}
Comparing with Eq.~\eqref{Eq:term_equivalence}, we find that we can compute the desired left-hand side
and thus the individual terms in Eq.~\eqref{Eq:gradient_decomp} from these quantities, since
\begin{equation}
  \bra{\psi}\G^{\dagger} \hat{Q} A  \ket{\psi} +\hc = 2 (p_0 \tilde{E}_0 - p_1 \tilde{E}_1) .
\end{equation}
Note that the measurement on the ancilla is not a typical conditional measurement with limited success probability: either result contributes to the final estimate.

Overall, this approach requires that we can apply the gate $\G$, as well the unitaries $A_k$ from the derivative decomposition in Eq.~\eqref{Eq:deriv_decomp}, controlled by an ancilla.
Altogether, we need to estimate $2 K$ expectation values and $2 K$ probabilities, and with Eq.~\eqref{Eq:decomp} $K$ can always be chosen as $2$.
The decomposition of $\partial_\mu \G$ into a linear combination of unitaries~$A_k$ needs to be found,
but this is easy for few qubit gates and has to be done only once.

Note that the idea of decomposing gates into ``\textit{classical} linear combinations of unitaries'' has been brought forward in Ref.~\cite{schuld2018circuit}, where $\hat{Q}$ had the special form of a $\sigma_z$ observable, which allowed the authors to evaluate expectations via overlaps of quantum states. Here we added the well-known strategy of \textit{coherent} linear combinations of unitaries \cite{childs12} to generalize the idea to any observable.

\section{Gradients of continuous-variable circuits}

We now turn to continuous-variable (CV) quantum computing architectures. Continuous-variable systems~\cite{weedbrook2012gaussian}
differ from discrete systems in that the generators of the gates typically have infinitely many unique eigenvalues,
or even a continuum of them.
Despite this, we can still find a version of the parameter-shift differentiation recipe which works for
Gaussian gates in CV variational circuits if the gate is only followed by Gaussian operations, and if the observable is a low-degree polynomial in the quadratures.
The derivation is based on the fact that in this case the effect of a Gaussian gate, albeit commonly represented by an infinite-dimensional matrix in the Schr\"{o}dinger picture, can be captured by a finite-dimensional matrix in the Heisenberg picture.

As in Sec.~\ref{sec:discrete}, the task is to compute $\partial_{\mu} \vc$.
In the Heisenberg picture, instead of evolving the state forward in time with the gates in the circuit, the final observable is evolved `backwards' in time with the adjoint gates.
We consider observables $\hat{B}$ that are polynomials of the quadrature operators~$\x_i$,~$\p_i$
(such as $\x_1\p_1\x_2$ or $\x_1^4 \p_2^3 +2\x_1$).
By linearity, it is sufficient to understand differentiation of the individual monomials.

For an $n$-mode system, we introduce the infinite-dimensional vector of quadrature monomials,
\begin{equation}
\hat{C} \coloneqq (\I, \x_1, \p_1, \x_2, \p_2, \ldots, \x_n, \p_n, \x_1^2, \x_1 \p_1, \ldots),
\end{equation}
sorted by their degree,
in terms of which we will expand the observables.

\subsection{CV gates in the Heisenberg picture}

Let us consider the Heisenberg-picture action $\G^{\dagger} \hat{C}_j \G$ of a gate $\G$ on a monomial $\hat{C}_j \in \hat{C}$.
This conjugation acts as a linear transformation $\Omega^\G$ on $ \hat{C}$, i.e.,
\begin{equation}
 \Omega^{\G}[\hat{C}_j] \coloneqq \G^\dagger \hat{C}_j \G = \sum_{i} M^{\G}_{ij} \hat{C}_i,
\end{equation}
where $M_{ij}^{\G} = M_{ij}^{\G}(\mu)$ are the  elements of a real matrix $M^{\G}$ that depends on the gate parameter.
Subsequent conjugations correspond to multiplying the matrices together:
\begin{equation}
  \Omega^{U}[\Omega^V[\hat{C}_k]] = \Omega^{U} [V^\dagger \hat{C}_k V] = \sum_{ij} M^{U}_{ij} M^{V}_{jk} \hat{C}_i.
\end{equation}

Suppose now that the gate $\G$ is Gaussian.
Conjugation by a Gaussian gate does not increase the degree of a polynomial.
This means that $\G$ will map the subspace of the zeroth- and first-degree monomials
spanned by $\hat{D} \coloneqq (\I, \x_1, \p_1, \x_2, \p_2, \ldots, \x_n, \p_n)$ into itself,
\begin{equation}
 \Omega^\G[\hat{D}_j] = \sum_{i=0}^{2n} M^{\G}_{ij} \hat{D}_i.
 \label{Eq:omega_gaussian}
\end{equation}
For observables that are higher-degree polynomials of the quadratures,
we can use the fact that $\Omega^\G$ is a unitary conjugation,
and that the higher-degree monomials can be expressed as
products of the lower-degree ones in~$\hat{D}$:
\begin{align}
  \label{eq:higher_degree_obs}
 \Omega^\G[\hat{D}_i\hat{D}_j]
 & = \mathcal{G}^\dagger \hat{D}_i \hat{D}_j \mathcal{G}, \\
 & = \mathcal{G}^\dagger \hat{D}_i \mathcal{G} \mathcal{G}^\dagger\hat{D}_j \mathcal{G}, \\
 & = \Omega^\G[\hat{D}_i] \Omega^\G[\hat{D}_j].
\end{align}
Hence we may represent any $n$-mode Gaussian gate~$\G$ as a $(2n+1) \times (2n+1)$ matrix in the Heisenberg picture.

We can now compute the derivatives $\partial_{\mu} \vc$
using the derivatives of the matrix~$M^{\G}(\mu)$.
It turns out that like the derivatives of the finite-dimensional gates in Sec.~\ref{sec:discrete},
$\partial_{\mu}M^{\G}$ can be often decomposed into a finite linear combination of matrices from the same class as~$M^{\G}$.
In fact, the derivatives of all gates from a universal Gaussian gate set can be decomposed to just two terms,
so derivative computations in this setting have the same complexity as in the qubit case.
We summarize the derivatives of important Gaussian gates in Table~\ref{Tbl:cv_gate_derivatives}.

\begin{table*}[t]
\def\arraystretch{1.2}
\begin{tabular*}{\textwidth}{l@{\extracolsep{\fill}}l@{\extracolsep{\fill}}l}
Gate~$\G$ & Heisenberg representation~$M^{\G}$  & Partial derivatives of~$M^{\G}$\\   \hline \hline
\cell{Phase rotation \\ $R(\phi)$}
&
$M^R(\phi) = \begin{pmatrix}
1 & 0          & 0 \\
0 & \cos\phi & -\sin\phi \\
0 & \sin\phi & \cos\phi
\end{pmatrix}$
&
$\partial_{\phi} \; M^R(\phi) =
\frac{1}{2} \big( M^R(\phi + \frac{\pi}{2})
-  M^R(\phi - \frac{\pi}{2}) \big)$\\   \hline
\cell{Displacement \\ $D(r, \phi)$}
&
$M^D(r, \phi) = \begin{pmatrix}
1 & 0 & 0 \\
2r \cos \phi & 1 & 0  \\
2r \sin \phi & 0 & 1 \\
\end{pmatrix}$
&
\cell{$\partial_{r} M^D(r, \phi) = \frac{1}{2s}\big( M^D(r+s,\phi) - M^D(r-s,\phi) \big), \; s \in \R$\\
$\partial_{\phi} M^D(r, \phi) = \frac{1}{2} \big( M^D(r,\phi+\frac{\pi}{2}) - M^D(r,\phi-\frac{\pi}{2}) \big)$
}\\  \hline
\cell{Squeezing\footnote{A more general version of the squeezing gate $\tilde{S}(r,\phi)$ also contains a parameter $\phi$ which defines the angle of the squeezing, and $S(r) = \tilde{S}(r,0)$. This two-parameter gate can be broken down into a product of single-parameter gates: $\tilde{S}(r, \phi) = R(\frac{\phi}{2})S(r)R(-\frac{\phi}{2})$.} \\ $S(r)$}
&
$M^S(r) = \begin{pmatrix}
        1 & 0 & 0 \\
        0 & \e^{-r}& 0 \\
        0 & 0 & \e^{r}
        \end{pmatrix}$
&
\cell{$\partial_{r} M^S(r) =   \frac{1}{2\sinh(s)} \big( M^S(r + s) -  M^S(r -s) \big), \; s \in \R$\\
}
\\ \hline
\cell{Beamsplitter\\ $B(\theta, \phi)$}
& \cell{
$M^B(\theta, \phi) = \begin{pmatrix}
            1 & 0 & 0 & 0 & 0\\
            0 & \cos\theta & 0 & -\alpha & -\beta \\
            0 & 0 & \cos\theta & \beta & -\alpha\\
            0 & \alpha & -\beta & \cos\theta & 0\\
            0 & \beta & \alpha & 0 & \cos\theta
        \end{pmatrix} $   }
&
\cell{$\partial_{\theta} M^B(\theta, \phi) =  \frac{1}{2} \big( M^B(\theta + \frac{\pi}{2}, \phi) - M^B(\theta - \frac{\pi}{2}, \phi) \big)$ \\
$\partial_{\phi} M^B(\theta, \phi) =  \frac{1}{2} \big( M^B(\theta, \phi+ \frac{\pi}{2}) -  M^B(\theta, \phi - \frac{\pi}{2}) \big)$\\
\\
$\alpha = \cos\phi\sin\theta, \;\;  \beta = \sin\phi\sin\theta$ 
}
\\   \hline \hline
\end{tabular*}
\caption{Parameter shift rules for the partial derivatives of important Gaussian gates.
  Every Gaussian gate can be decomposed into this universal gate set.
  We use the gate definitions laid out in the Strawberry Fields documentation~\cite{killoran2018strawberry} with $\hbar=2$. 
  All parameters are real-valued.
  Single-mode gates have been expanded using the set $(\I,\x,\p)$, whereas the
  two-mode beamsplitter has been expanded using the set
  $(\I, \x_a, \p_a, \x_b, \p_b)$. More derivative rules can be found in the PennyLane \cite{bergholm2018pennylane} documentation (https://pennylane.readthedocs.io)}
\label{Tbl:cv_gate_derivatives}
\end{table*}

As an example, we consider the single-mode squeezing gate with zero phase $S(r, \phi=0)$, which is represented by
\begin{equation}\label{eq:squeezinggate}
  M^S(r) =
	\begin{pmatrix}
	1 & 0 & 0 \\
	0 & e^{-r} & 0 \\
	0 & 0 & e^r
	\end{pmatrix}.
\end{equation}
Its derivative is given by
\begin{equation}
	\partial_{r} \; M^S(r) =
	\begin{pmatrix}0 & 0 & 0 \\
	0 & -e^{-r} & 0 \\
	0 & 0 & e^r
	\end{pmatrix}.
\end{equation}
The derivative itself is not a Heisenberg representation of a squeezing gate, but we can decompose it into a linear combination of such representations, namely
\begin{equation}
\partial_r \; M^S(r) =  \tfrac{1}{2\sinh(s)}\left(M^S(r + s) - M^S(r - s)\right),
\end{equation}
where $s$ is a fixed but arbitrary nonzero real number.
Hence,
\begin{align}
  \notag
  \partial_{r} \left[S(r)^\dagger \hat{B}_j S(r)\right]
  =& \tfrac{1}{2\sinh(s)} \big(S(r + s)^{\dagger} \hat{B}_j S(r + s) \\
  & -S(r - s)^{\dagger} \hat{B}_j S(r - s)\big)
\end{align}
for $j \in \{0,1,2\}$.

\subsection{Differentiating CV circuits}

Again we split the gate sequence into three pieces,
$U(\theta) = V \G(\mu) W$.
For simplicity, let us at first assume that our observable
is a first-degree polynomial in the quadrature operators, and thus
can be expanded as
$\hat{B} = \sum_i b_i \hat{D}_i$.
As shown in the previous section, for Gaussian gates the
Heisenberg-picture matrix~$M$ is block-diagonal, and maps
from the space spanned by $\hat{D}$ onto itself.
Thus, if $\G$ is Gaussian and $V$~consists of Gaussian gates only,
we may write
\begin{align}
  \vc(\theta) &= \bra{0} U^{\dagger}(\theta) \hat{B} U(\theta) \ket{0},\\
  &=
  \sum_{ijk} \: \bra{0} W^\dagger \hat{D}_k W \ket{0} M^\G_{kj}(\mu) M^V_{ji} b_i,
\end{align}
where $\ket{0}$ denotes the vacuum state.
Now the derivative is simply
\begin{align}
  \partial_\mu \vc(\theta)
  &=
  \sum_{ijk} \: \bra{0} W^\dagger \hat{B}_k W \ket{0} (\partial_\mu M^\G)_{kj} M^V_{ji} b_i.
\end{align}
If $\partial_\mu M^\G$ can be expressed as a linear combination $\sum_i \gamma_i M^\G(\mu+s_i)$ with $\gamma_i, s_i \in \mathbb{R}$,
by linearity we may express $\partial_\mu \vc$ using the same linear combination,
$\partial_\mu \vc = \sum_i \gamma_i \vc(\mu +s_i)$. This is the parameter shift rule for CV quantum computing.

What about the subcircuit $W$ that appears before the gate that we differentiate?
For the purposes of differentiating the gate $\G$, this subcircuit can be arbitrary,
since the above differentiation recipe does not depend on the properties of the matrix~$M^W$.
The above recipe works as long as no non-Gaussian gates are between $\G$ and the observable~$\hat{B}$.

With observables $\hat{B}$ that are higher-degree polynomials of the quadratures,
we can use the property in Eq.~\eqref{eq:higher_degree_obs}
to compute the derivative using the product rule:
\begin{align}
  \partial_\mu\left(\Omega^\G[\hat{B}_i\hat{B}_j] \right)
  & = \partial_\mu\left(\Omega^\G[\hat{B}_i] \Omega^\G[\hat{B}_j]\right),  \\
  & = \partial_\mu\Omega^\G [\hat{B}_i] \; \Omega^\G[\hat{B}_j]
  +\Omega^\G[\hat{B}_i] \; \partial_\mu \Omega^\G[\hat{B}_j].
  \notag
\end{align}

\subsection{Non-Gaussian transformations}

For the above decomposition strategy to work efficiently, the subcircuit $V$ must be Gaussian.
In the case that $V$ is non-Gaussian, it will generally increase the degree of the final observable, i.e., $V^\dagger \hat{B} V$ will be higher degree than $\hat{B}$.
For example, the cubic phase gate $V(\gamma) = \e^{i \gamma \hat{x}^3}$ carries out the transformations
\begin{align}
 V^\dagger(\gamma)\x V(\gamma) = &~\x, \\
 V^\dagger(\gamma)\p V(\gamma) = &~\p + \gamma x^2.
\end{align}
In this case, we will have to consider a higher-dimensional subspace (tracking both the linear and the quadratic terms).
If the subcircuit $V$ contains multiple non-Gaussian gates, each one can raise the degree of the observable.
Thus, the matrices considered in the Heisenberg representation can become large depending on both the quantity and the character of non-Gaussian gates in the subcircuit $V$.
Finding analytic derivative decompositions of circuits containing non-Gaussian gates is more challenging, but not strictly ruled out by complexity arguments. Specifically, in the case where there are only logarithmically few non-Gaussian gates, and each of those gates only raises the degree of quadrature polynomials by a bounded amount, there is still the possibility to efficiently decompose a gradient of an expectation value into a polynomial number of component expectation values.

\section{Conclusion}

We present several hardware-compatible strategies to evaluate the derivatives of quantum expectation values from the output of variational quantum circuits.
In many cases of qubit-based quantum computing the derivatives can be computed with a simple parameter shift rule, using the variational architecture of the original quantum circuit.
In all other cases it is possible to do the same by using an ancilla and a decomposition of the ``derivative of a gate''.
For continuous-variable architectures we show that, as long as the parameter we differentiate with respect to feeds into a Gaussian gate that is only followed by Gaussian operations, a close relative to the parameter shift rule can be applied.
We leave the case of non-Gaussian circuits as an open direction for future research.

\bibliography{references}

\end{document}